\documentclass[12pt,preprint]{aastex}

\slugcomment{Submitted to the Astrophysical Journal Letters}

\shorttitle{The first subdwarf with spectral type sdM8.0}
\shortauthors{Lepine, Shara, \& Rich}

\begin{document}

\title{Discovery of an ultra-cool subdwarf: LSR1425+7102, first star
 with spectral type sdM8.0}

\author{S\'ebastien L\'epine\altaffilmark{1,2,3}, Michael
M. Shara\altaffilmark{1}, and R. Michael Rich\altaffilmark{4}}
\altaffiltext{1}{Department of Astrophysics, Division of Physical
Sciences, American Museum of Natural History, Central Park West at
79th Street, New York, NY 10024, USA, lepine@amnh.org, shara@amnh.org}
\altaffiltext{2}{Kalbfleich research fellow}
\altaffiltext{3}{Visiting Astronomer, KPNO}
\altaffiltext{4}{Department of Physics and Astronomy, University of
California at Los Angeles, Los Angeles, CA 90095, USA,
rmr@astro.ucla.edu}

\begin{abstract}
We report the discovery of the coolest subdwarf reported to date. The
star LSR1425+7102 was discovered in our survey for faint high proper
motion stars in the northern sky. Follow-up spectroscopy revealed the
star to be a very red object with the characteristic signature of M
subdwarfs: strong CaH bands but relatively weaker TiO bands. The CaH
molecular band at $\approx\lambda$6900\AA\ is particularly strong. By
extrapolating the empirical relationship between the strength of the
CaH molecular band and the subdwarf subtype, we conclude that
LSR1425+7102 is the first star to be discovered with spectral type
sdM8.0.
\end{abstract}


\section{Introduction}

Metal-poor stars are rare in the neighborhood of the Sun, and the
current sample of local subdwarfs is very limited. In particular, very
few low-mass subdwarfs are known and, as a result, the bottom of the
main sequence is very poorly constrained for metal poor stars. While
we now have a well populated sequence of known low-mass,
solar-metallicity red dwarfs extending well into the brown dwarf
regime \citep{D02,H02}, we are still seriously lacking in nearby
prototypes for bottom of the main sequence subdwarfs. The search for
low-mass, low-luminosity subdwarfs is hampered not only by the fact
that metal-poor stars are rare and low-mass subdwarfs are intrinsically
faint, but also because late-type M subdwarfs do not show
exceptionally red colors as ultra-cool M dwarfs do.

M subdwarfs, however, can be readily distinguished from M dwarfs
because of their very characteristic spectral signature. The optical
spectra of M dwarfs and subdwarfs are dominated by molecular
absorption bands, mainly of oxides (TiO, VO) and hydrides (CaH,
MgH, FeH). While the oxide bands dominate in M dwarf spectra, they are
markedly weaker in subdwarfs, where the hydride bands dominate
\citep{EG65}. The relative strengths of the TiO and CaH bands can thus
be used as an objective, spectroscopic criterion in the identification
and classification of M subdwarfs \citep{MM78}. A quantitative scheme
for the spectroscopic classification of M dwarfs and subdwarfs was
developed by Gizis (1997a, hereafter G97a) based on the strength of
the CaH and TiO bands around 7000\AA. Stars are objectively separated
into three classes, the M dwarfs (M V), the M subdwarfs (sdM), and the
extreme M subdwarfs (esdM), based on spectroscopic criteria. The
metallicity range for each of the three classes corresponds roughly to
[Fe/H]$\approx0.0$ for M dwarfs, [Fe/H]$\approx-1.2\pm0.3$ for sdM,
and [Fe/H]$\approx-2.0\pm0.5$ for esdM. Observations of M subdwarfs
that are companions to F and G subdwarfs of known metallicities have
confirmed the consistency of this scale \citep{GR97b}.

While there are now numerous examples of ultra-cool M dwarfs
\citep{KHS95,KHI97}, only a handful of subdwarfs with spectral subtype
6.0 or later are currently known. At the time the classification
system of G97a was introduced, only one ultra-cool sdM was known:
the high proper motion star LHS377 (spectral type sdM7.0). The
coolest esdM then known was LHS1742a (esdM5.5). A
follow-up spectroscopic survey of faint high proper motion stars from
the LHS catalog \citep{L79} for which no spectra had been obtained
yet, revealed the existence of two more very cool subdwarfs: LHS1035
(sdM6.0), and LHS1135 (sdM6.5) \citep{GR97a}. That survey also
revealed LHS1826 to be the first extreme subdwarf with spectral type
esdM6.0. Follow-up spectroscopy of new high proper motion stars in the
southern sky by \citet{SSSIM99} lead to the discovery of the coolest
esdM known to date, APMPM J0559--2903 (esdM7.0). Another ultra-cool
sdM was found recently by \citet{LRS03} from a follow-up survey of
newly discovered high proper motion stars in the northern sky:
LSR2036+5059 (sdM7.5).

In this paper, we report the discovery of the fifth known subdwarf
with spectral subtype 6.0 or later, and the coolest reported to
date. This is the faint high proper motion star LSR1425+7102, to which
we assign a spectral type sdM8.0.

\section{Proper Motion Discovery and Photometry}

The high proper motion star LSR1425+7102 was discovered as part of
our new search for high proper motion stars in the northern sky using
the Digitized Sky Survey \citep{LRS02}, performed as a part of the
NStars initiative. The star was found in a relatively low density
field at a moderate galactic latitude (b=+44). The star is not
recorded in the Luyten catalogs of high proper motion stars, and a
search on Simbad (http://simbad.u-strasbg.fr/Simbad) around that
location on the sky yielded no results, confirming that this high
proper motion star is being reported here for the first time. The
Digitized Sky Survey discovery fields are presented in Figure 1. The
left panel shows the 1955 POSS-I red (103aE + red plexiglass) image of
a $4.25\arcmin\times4.25\arcmin$ field centered on the position of the
star at epoch 2000.0. The right hand side shows the 1991 POSS-II red
(IIIaF + RG 610) image of the same field. The motion of LSR1425+7102
over the 36 years period is very apparent. The star has a proper
motion $\mu_{RA}=-0.61\arcsec$yr$^{-1}$ and
$\mu_{DEC}=-0.17\arcsec$yr$^{-1}$. The coordinates of LSR1425+7102 are
listed in Table 1 along with the various parameters mentioned in this
paper.

We have found this star to match the star listed as 1610--0102604 in
the USNO-1B catalog. The USNO-1B catalog gives the blue, red, and near
infra-red photographic magnitudes from the POSS-II plates. This gives
LSR1425+7102 photographic magnitudes b=20.8, r=18.6, and i=16.2,
making it a fairly red object, consistent with an M dwarf or
subdwarf. We also found LSR1425+7102 to match the source 2MASS
1425050+710209, which is listed in the {\it 2MASS Point Source Catalog
Second Incremental Release}. The 2MASS magnitudes yield an infrared
color J--K=0.5, which is consistent with the star being a
subdwarf. Observations show that all M dwarfs have J--K$>$0.7, and
only M subdwarfs are found with J--K$<$0.7 \citep{LAH98}. 

\section{Spectroscopy}

The star LSR1425+7102 was observed on the night of 19 May 2002, at the
4-m Mayall telescope of the Kitt Peak National Observatory. A spectrum
of the star was obtained with the R-C spectrograph equipped with the
LB1A CCD camera, and mounted at the Cassegrain focus. We used the
BL181 disperser (316 l/mm, blazed at 7500\AA), with the OG530 order
blocking filter. The star was imaged through a 1.5$\arcsec$ slit,
yielding a 5\AA\ spectral resolution. Standard spectral reduction was
performed with IRAF using the CCDPROC and SPECRED packages, including
removal of telluric features. Calibration was derived from
observations of the standard Feige 34 \citep{MG90}. Both the target
and the standard were observed at the smallest possible airmass
($<1.3$) and with the slit at the parallactic angle to minimize slit
loss due to atmospheric diffraction, providing excellent
spectrophotometric calibration.

We measured the radial velocity of the star from the centroids of the
\ion{K}{1} $\lambda\lambda$7665,7699, \ion{Na}{2}
$\lambda\lambda$8183,8195, and \ion{Ca}{2} $\lambda\lambda$8540,8660
atomic lines. After correction for the earth's motion in space, and
accounting for uncertainties, we find an estimated heliocentric radial
velocity v$_{hel}=-60\pm20$km s$^{-1}$.

\section{Assignment of spectral type sdM8.0}

A spectrum of LSR1425+7102 is plotted in Figure 2, where it is
compared to the spectrum of LSR2000+3057, a known M6.0 V dwarf
\citep{LRS03}, observed at a similar spectral resolution. The general
slope of the spectral energy distribution in the 6000-9000\AA\ range
for both stars are very similar. Note also the great similarity in the
\ion{Na}{1} doublet profiles. The difference between the two stars
clearly is in the relative strength of the CaH and TiO molecular
bands. The TiO bands dominate in the M6.0V star, but they are much
weaker in LSR1425+7102, in which the CaH bands are marginally
stronger. It is remarkable that the TiO band at $\lambda$8430 is very
weak in LSR1425+7102. The star thus clearly shows the classic
signature of a subdwarf, although a very cool one.

We quantify the behavior of the CaH and TiO bands by calculating
values for the CaH1, CaH2, CaH3, and TiO5 indices defined in
\citet{RHG95}, and which serve as the main classification criteria for
subdwarfs (see G97a). We find for LSR1425+7102 the following values:
CaH1=0.309, CaH2=0.200, CaH3=0.306, and TiO5=0.307. To our knowledge,
the very low value of the CaH3 index is the lowest ever measured in a
low-mass star, a record previously held by the esdM7.0 star APMPM
J0559--2903 \citep{SSSIM99}, for which CaH3=0.32 had been
measured. This clearly makes LSR1425+7102 one the coolest subdwarfs to
date.

The separation between the sdM and esdM classes as defined by
G97a is based on the position of the star in the CaH2/TiO5
diagram. Under this system, LSR1425+7102 clearly falls in the range of
sdM stars (see Figure 1 in G97a). The mean value of two separate
relationships (respectively based on the CaH2 and CaH3 indices) are
used to calculate the spectral subclass, rounded to the nearest half
integer. For LSR1425+7102, it yields a spectral type sdM8.0. These
spectral index to spectral subtype relationships were originally
defined only to spectral subtype 7.0, but it seems reasonable to
extrapolate them at least to spectral subtype 8.0, since the CaH band
does not appear to begin to saturate yet.

In Figure 3, we plot CaH2+CaH3 against TiO5 for an extended sample of
dwarfs, subdwarfs, and extreme subdwarfs. The points in the diagram
represent all stars whose spectroscopic indices were published in
G97a, \cite{GR97a}, \citet{CR02}, and \citet{LRS02}, plus the esdM7.0
star APMPM J0559--2903 \citep{SSSIM99}, the esdM5.0 star LP 382-40
\citep{GR99} and LSR1425+7102 (this paper). All sdM and esdM with
published values of CaH2, CaH3, and TiO5 are thus represented
here. Many more nearby M dwarfs ($>$1600) have also had their spectral
indices measured \citep{RHG95,HGR96}, but are not plotted on this
graph. All the known ultra-cool sdM and esdM (spectral subtype 7.0 or
later) are circled and labeled by name. The distribution has a
significant gap in the vicinity of [CaH2+CaH3,TiO5]=[0.5,0.7], but the
region occupied by the three known late-type sdM stars (including
ours) suggest that the void may well be due to the small number of
such stars actually discovered, rather than indicating a physical
boundary in their spectroscopic properties.

\section{Discussion and Conclusions}


Because LSR1425+7102 is an sdM, its metallicity is
[Fe/H]$\approx-1.2\pm0.3$. Based on the evolutionary models
of \citet{BCAH97} for low-mass, metal-poor stars, a subdwarf with
[Fe/H]$=-1.5$, J--K$=0.5$ would have a mass M$\simeq$0.095M$_{\sun}$,
while a subdwarf with [Fe/H]$=-1.0$, J--K$=0.5$ would have a mass
M$\simeq$0.088M$_{\sun}$. The magnitude and color relationships
presented in \citet{BCAH97} are based on the ``NextGen'' model
atmospheres, which does not include the formation of atmospheric
dust. This model has since been superseded by the ``DUSTY'' atmosphere
model \citet{CBAH00}, which yields significantly different magnitudes
and colors for low-mass stars with solar abundances. It is very
possible that the use of the ``DUSTY'' model for metal-poor stars also
yields slightly different results. Because of the uncertainty, we
adopt for LSR1425+7102 a conservative mass estimate of
M=0.09$\pm$0.01M$_{\sun}$.


We again use the models of \citet{BCAH97} to estimate the distance to
LSR1425+7102. Their theoretical, absolute magnitudes for a
metal-poor 0.09M$_{\sun}$ star with [Fe/H]$=-1.3$ are J=10.74, H=10.31,
and K=10.27. This suggests a distance modulus $\simeq$4.1 for
LSR1425+7102. Accounting for possible uncertainties in the models, we
adopt a conservative distance estimate of 65$\pm15$pc. At that
distance, the proper motion and radial velocity yield components of
the space motion U$=-65\pm22$, V$=-177\pm35$, and W$=+64\pm27$, where
U is in the direction of the galactic center, V is towards the
direction of galactic rotation, and W towards the galactic north
pole. These values make LSR1425+7102 a probable halo member.


Deep photometry of the globular cluster Messier 4
\citep{RBFGHIKLRSSS02} has recently showed that the bottom of its main
sequence extends to significantly fainter luminosities and cooler
temperatures than suggested by the current sample of field sdM/esdM
stars. Based of these results, \citet{RBFGHIKLRSSS02} have postulated
the existence of very cool subdwarfs in the solar neighborhood,
although they argued that these stars should be extremely rare (1 star
in 30,000). Our discovery of LSR1425+7102 brings support to this
hypothesis. The observation that at least one sdM8.0 star with halo
kinematics does exist within 100pc of the Sun strongly indicates that
similar objects must exist in large numbers throughout the
Galaxy. Furthermore, the extended Messier 4 main sequence suggests
that subdwarfs even cooler than LSR1425+7102 should also be found in
the neighborhood of the Sun.

The current sample of spectroscopically identified cool and ultra-cool
subdwarfs is thus still clearly lacking in representative
objects. More efforts should be devoted to the search and
spectroscopic identification of cool sdM and esdM stars, so that their
contribution to the Galactic mass can be evaluated. A systematic spectral
classification of faint stars with large proper motions should result
in the recovery of more cool and ultra-cool subdwarfs. A more complete
sample of sdM and esdM over a large range of temperatures and
abundances would provide much needed constraints for current and
future atmospheric models of low-mass, metal-poor stars.

\acknowledgments

This research program is being supported by NSF grant AST-0087313 at
the American Museum of Natural History, as part of the NSTARS initiative.

\newpage

\begin{deluxetable}{lrl}
\tabletypesize{\scriptsize}
\tablecolumns{3} 
\tablewidth{0pt} 
\tablecaption{Basic Data for LSR1425+7102} 
\tablehead{Datum & Value & Units}
\startdata 
RA (2000.0) & 14:25:04.81 & h:m:s\\
DEC (2000.0) & +71:02:10.4 & d:m:s\\
$\mu$       & 0.635 & $\arcsec$ yr$^{-1}$\\
pma         & 74.7 & $\degr$\\
v$_{hel}$   & -65$\pm$20 & km s$^{-1}$\\
b     &   20.8$\pm$0.5\tablenotemark{1} & mag\\
r     &   18.6$\pm$0.5\tablenotemark{1} & mag\\
i     &   16.2$\pm$0.5\tablenotemark{1} & mag\\
J     & 14.83$\pm$0.04\tablenotemark{2} & mag\\
H     & 14.43$\pm$0.06\tablenotemark{2} & mag\\
K$_s$ & 14.34$\pm$0.10\tablenotemark{2} & mag\\
CaH1  & 0.309\tablenotemark{3} & \\
CaH2  & 0.200\tablenotemark{3} & \\
CaH3  & 0.306\tablenotemark{3} & \\
TiO5  & 0.307\tablenotemark{3} & \\
Spectral Type & sdM8.0& \\
Distance & 65$\pm$15 & pc \\
$U$ &  -65$\pm$22 & km s$^{-1}$\\
$V$ & -177$\pm$35 & km s$^{-1}$\\
$W$ &   64$\pm$27 & km s$^{-1}$
\enddata
\tablenotetext{1}{Photographic B, R, and I magnitudes from USNO B-1.0 catalog.}
\tablenotetext{2}{Infrared magnitudes from the 2MASS survey.}
\tablenotetext{3}{Spectroscopic indices (see text).}
\end{deluxetable} 

\newpage

\begin{figure}
\plotone{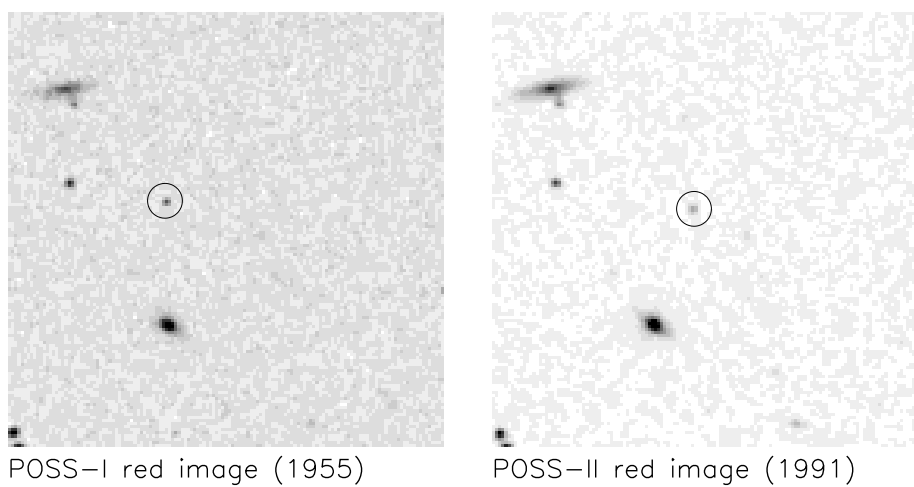}
\caption{\label{fig1} The new high proper motion star
LSR1425+7102. Left: red plate of the first epoch Palomar Sky Survey,
obtained in 1955. Right: red plate of the second epoch Palomar Sky
Survey, obtained in 1991. All the fields are $4.0\arcmin$ on the side,
with north up and east left. Circles are drawn centered on the
location of LSR1826+3014 at each epoch. The star is moving with a
proper motion $\mu=0.635\arcsec$yr$^{-1}$.}
\end{figure}

\begin{figure}
\plotone{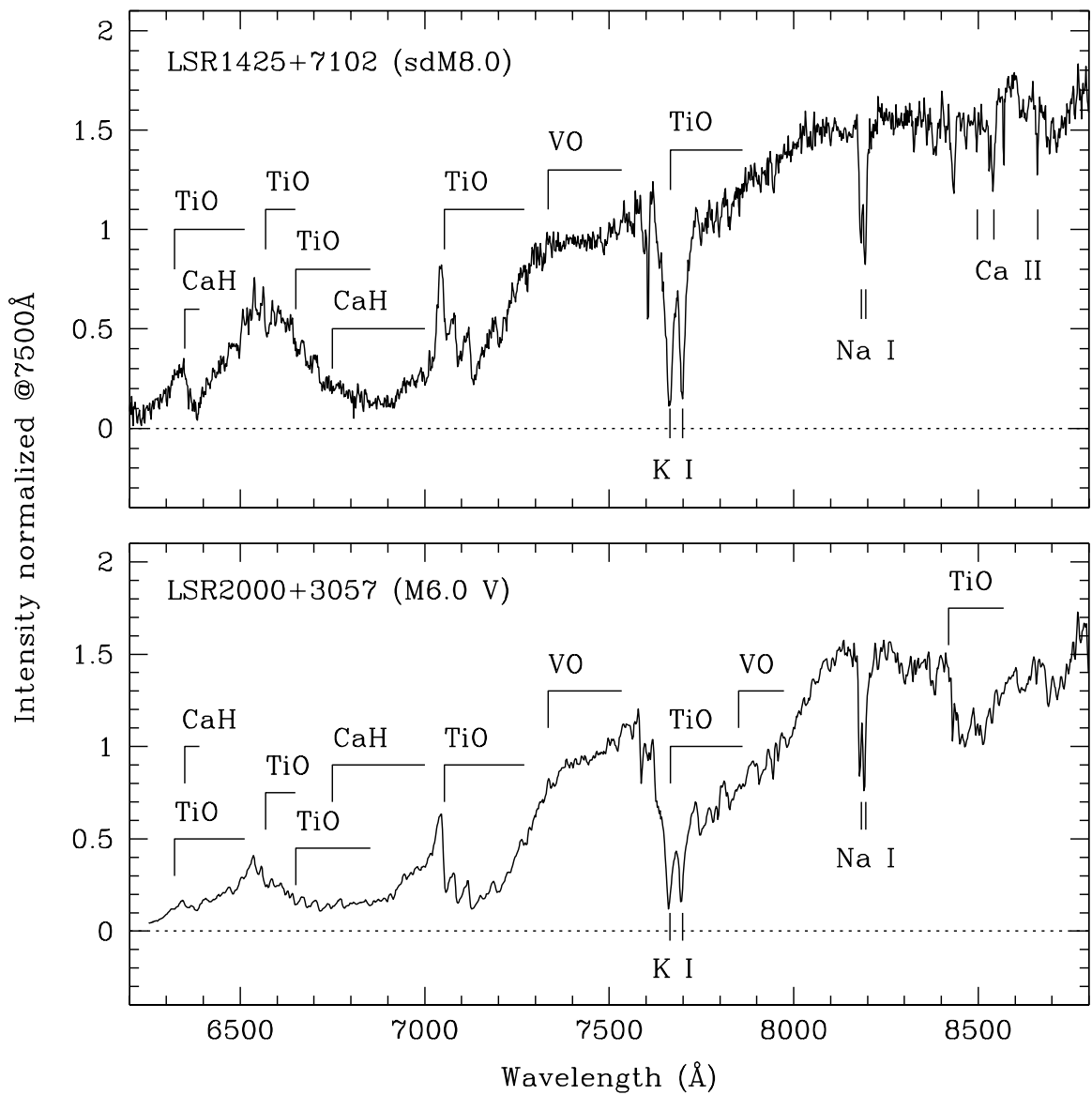}
\caption{\label{fig2} Optical spectrum of the sdM8.0 high proper
motion star LSR1425+7102 obtained with the R-C spectrograph on the 4m
Mayall Telescope at KPNO (top). Shown for comparison is a spectrum of
the M6.0 V star LSR2000+3057 (bottom). The very deep CaH band and the
very red spectral energy distribution point to a very late spectral
type for LSR425+7102, but the TiO bands are conspicuously weak as
compared with those of the M6.0 V star. Thus LSR1425+7102 is more
consistent with a low metallicity M subdwarf. The depth of the CaH
bands suggests a spectral type sdM8.0, making LSR1425+7102 the coolest
subdwarf known, and the first of its subtype.}
\end{figure}

\begin{figure}
\plotone{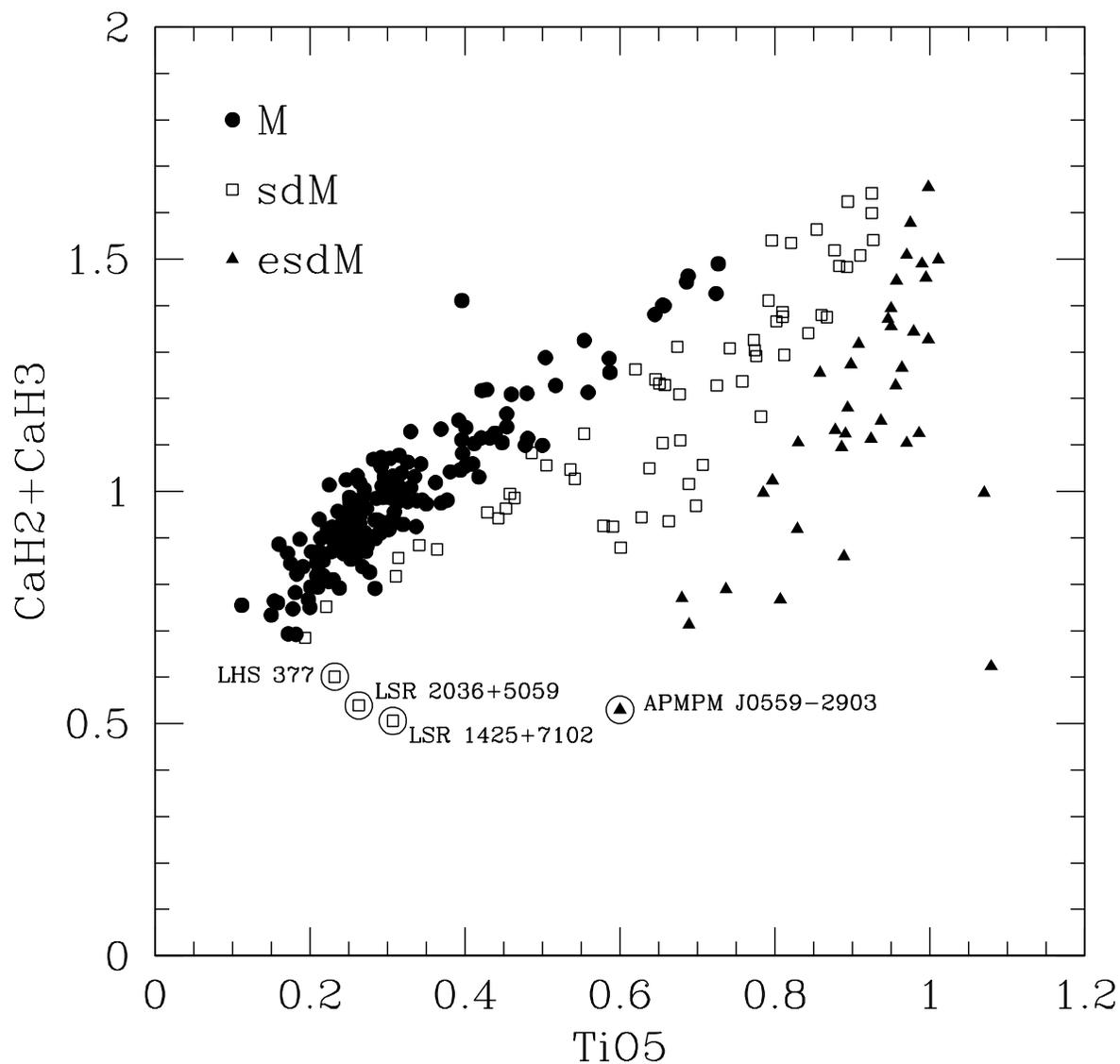}
\caption{\label{fig3} Distribution of M dwarfs, subdwarfs (sdM), and
  extreme subdwarfs (esdM) in the CaH2+CaH3 versus TiO5 spectral
  indices diagram (see text). Early-type stars are on the upper right
  of the distribution, late-type stars on the lower left, and the
  different metallicity classes are well segregated. This distribution
  is complete for sdM and esdM with published values of CaH2, CaH3,
  and TiO5. Ultra cool sdM and esdM (subtype 7.0 and later) are
  circled and identified by name.}
\end{figure}


\end{document}